\documentclass[a4paper,11pt]{article}
\usepackage{pos}

\usepackage{cleveref}
\usepackage[olditem,oldenum]{paralist}
\usepackage{siunitx}
\usepackage{subcaption}

\title{Why interpretation matters for BSM searches:\\a case study with Heavy Neutral Leptons at ATLAS}
\ShortTitle{Why interpretation matters for BSM searches: a case study with HNLs at ATLAS}

\author*[a,b,c]{Jean-Loup Tastet}
\author[a]{Oleg Ruchayskiy}
\author[a,b]{Inar Timiryasov}

\affiliation[a]{Niels Bohr Institute, University of Copenhagen,\\
  Blegdamsvej 17, DK-2010, Copenhagen, Denmark}

\affiliation[b]{Institute of Physics, Laboratory for Particle Physics and Cosmology,\\
  École Polytechnique Fédérale de Lausanne, CH-1015 Lausanne, Switzerland}

\affiliation[c]{Instituto de Física Teórica, Universidad Autónoma de Madrid,\\
  Calle Nicolás Cabrera 13-15, Cantoblanco E-28049 Madrid, Spain}

\emailAdd{jean-loup.tastet@uam.es}
\emailAdd{inar.timiryasov@nbi.ku.dk}
\emailAdd{oleg.ruchayskiy@nbi.ku.dk}

\abstract{%
  Experiments searching for Heavy Neutral Leptons (HNLs) typically interpret their results within simplified models consisting of a single HNL coupled to a single lepton flavor. However, any model which aims to describe neutrino oscillations necessarily features more than one HNL, coupled to several flavors. As we show in this work, the reinterpretation of the results of experimental searches in terms of realistic models is a non-trivial task. We perform a detailed reinterpretation of the latest ATLAS search for prompt HNLs in $W$ decays within a minimal low-scale seesaw with two HNLs. We show that the exclusion limits obtained using the detailed reinterpretation can differ by \emph{several orders of magnitude} from the limits quoted for the simplified models. Hence naively comparing the mixing angles from a realistic model to the reported limits could lead to \emph{wrongly excluding} entire regions of parameter space! To overcome this issue without requiring experiments to report constraints on all possible HNL models, we propose a simple framework that allows one to easily and accurately reinterpret exclusion limits within closely-related models. We outline a number of concrete steps that can be taken by experiments to implement this method with minimal effort, and we discuss its applicability to other models of feebly interacting particles.
}

\FullConference{%
  *** The European Physical Society Conference on High Energy Physics (EPS-HEP2021), ***\\
  *** 26-30 July 2021 ***\\
  *** Online conference, jointly organized by Universität Hamburg and the research center DESY ***
}

\begin{document}
\maketitle

In this talk, based on ref.~\cite{Tastet:2021vwp}, we take the example of a search~\cite{Aad:2019kiz} at ATLAS for Heavy Neutral Leptons (HNLs) to demonstrate the importance of reinterpretation.
In \cref{sec:introduction} we briefly introduce the model, before discussing in \cref{sec:constraints} the constraints set by this ATLAS search on HNLs. In \cref{sec:parameter_space} we describe the parameter space of this model, and in \cref{sec:reinterpretation} our reinterpretation method and its findings. Finally, in \cref{sec:conclusion}, we discuss our experience and give some suggestions to experiments for improving the reinterpretability of their results.

\section{Introduction to the model}
\label{sec:introduction}

The Standard Model has some well-known observational shortcomings: it
\begin{inparaenum}[\it (a)]
    \item does not contain neutrino masses at the renormalizable level
    \item cannot explain the observed baryon asymmetry of the Universe (BAU)
    \item cannot explain dark matter.
\end{inparaenum}
One of the many possible solutions to these problems consists in adding to the Standard Model two or more right-handed neutrinos $N_I$ --- or \emph{Heavy Neutral Leptons} (see e.g.\ ref.~\cite{FIPs}).
As Standard Model singlets (i.e.\ completely neutral particles), they admit a Majorana mass term which, combined with the Yukawa interaction, produces a non-diagonal mass term after electroweak symmetry breaking. This leads to mixing between the neutrino flavor states $\nu_{\alpha}$ and the new heavy mass eigenstates $N_I$, which thus behave as heavy Majorana neutrinos with interactions suppressed by a small mixing angle $\Theta_{\alpha I}$:
\begin{equation*}
    \nu_\alpha \approx V_{\alpha i}^{\mathrm{PMNS}} \nu_i + \Theta_{\alpha I} N_I^c.
\end{equation*}

\section{ATLAS constraints on HNLs}
\label{sec:constraints}

HNLs have elicited a strong interest from the experimental community. Here we focus on a specific search~\cite{Aad:2019kiz} by the ATLAS collaboration, for HNLs in the mass range $M_N \in [5,50]\,\si{GeV}$, produced in $W$ decays and decaying promptly to the trilepton final states $e^{\pm} e^{\pm} \mu^{\mp}$ (electron channel) and $\mu^{\pm} \mu^{\pm} e^{\mp}$ (muon channel) plus missing transverse energy. Both channels have contributions from both lepton number conserving (LNC) and lepton number violating (LNV) processes.
Like most experiments, ATLAS has reported their limits for simplified models only, where a single Majorana HNL mixes with either the electron or muon neutrino, but not both. The LNC processes depend on both mixing angles, therefore their contribution was not included in this original interpretation.

\section{Parameter space of the model}
\label{sec:parameter_space}

The seesaw mechanism, being responsible for the generation of neutrino masses, relates the HNL masses and mixing angles to the measured neutrino oscillation parameters~\cite{NuFIT5.0}.
In addition, if HNLs have roughly the same interaction strengths and are within experimental reach, then it can be shown that their masses must be nearly degenerate \cite{Kersten:2007vk,Drewes:2019byd}.
In what follows, we will focus on a minimal seesaw model with only two nearly degenerate HNLs.

\paragraph{Constraints on the mixing angles}

\begin{figure}
    \centering
    \includegraphics[width=0.76\textwidth]{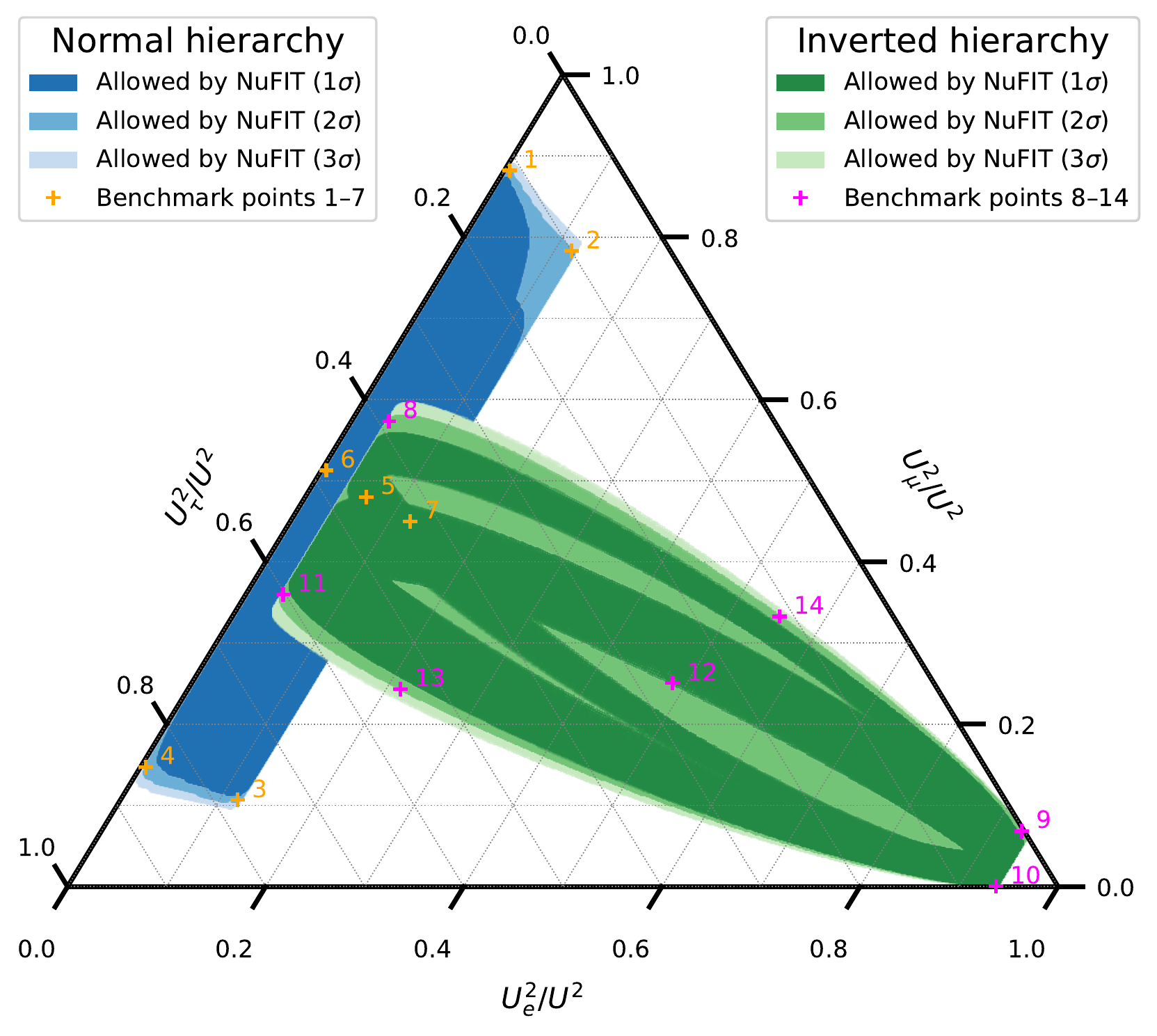}
    \caption{Allowed ratios of the three squared mixing angles $|\Theta_{eI}|^2$, $|\Theta_{\mu I}|^2$ and $|\Theta_{\tau I}|^2$.}
    \label{fig:ternary_plot}
\end{figure}

From the point of view of collider experiments, a pair of nearly degenerate HNLs will behave as a single particle. Combining this degeneracy with the seesaw formula and neutrino oscillation data~\cite{NuFIT5.0}, we obtain a constraint on the allowed ratios of squared mixing angles with the electron, muon and tau flavors, as shown in \cref{fig:ternary_plot} \cite{Drewes:2016jae,Caputo:2017pit}.
The original interpretation set constraints on only two points in this plane: the right (electron channel) and top (muon channel) vertices of this triangle, which can be seen to be incompatible with neutrino oscillation data within the model under consideration.\footnote{These constraints would be significantly relaxed by the addition of extra nearly-degenerate HNLs.}

\paragraph{HNL oscillations}

\begin{figure}
    \centering
    \begin{subfigure}[b]{0.33\textwidth}
        \centering
        {\hspace*{2em}\small($\delta M/\Gamma = \pi/10$)}\\
        \includegraphics[width=\linewidth]{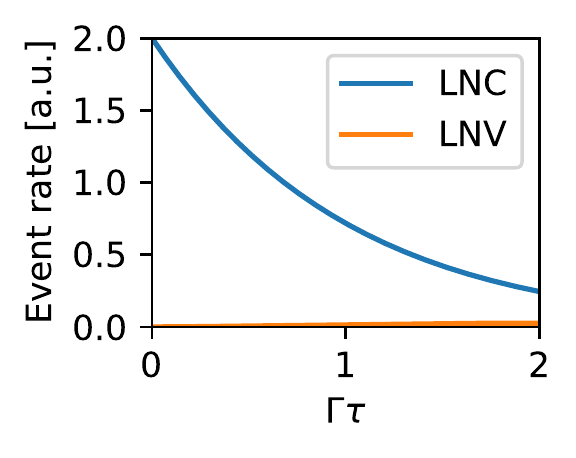}
        \caption{Dirac-like}
        \label{fig:osc_Dirac-like}
    \end{subfigure}\hfill%
        \begin{subfigure}[b]{0.33\textwidth}
        \centering
        {\hspace*{2em}\small($\delta M/\Gamma = \pi$)}\\
        \includegraphics[width=\linewidth]{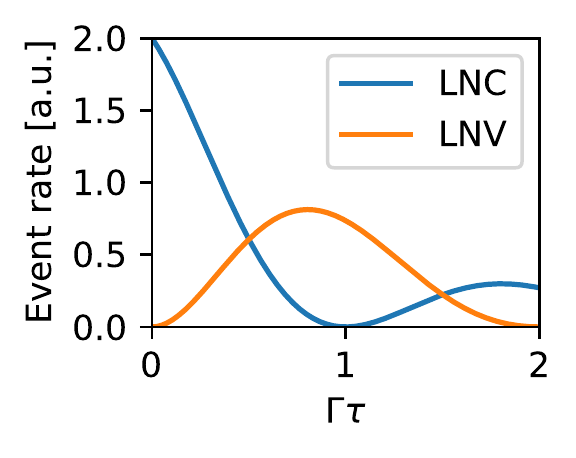}
        \caption{Visible oscillations}
    \end{subfigure}\hfill%
        \begin{subfigure}[b]{0.33\textwidth}
        \centering
        {\hspace*{2em}\small($\delta M/\Gamma = 10\pi$)}\\
        \includegraphics[width=\linewidth]{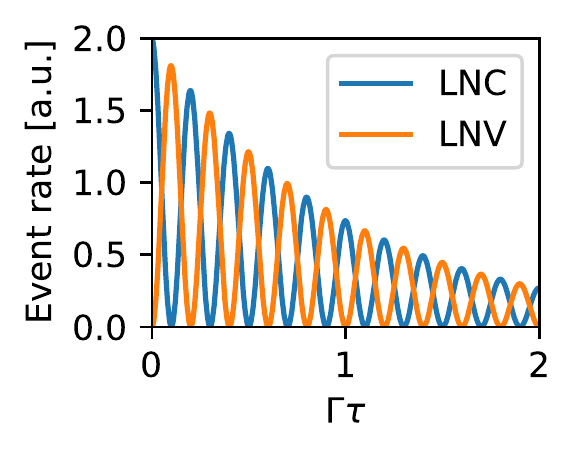}
        \caption{Majorana-like}
        \label{fig:osc_Majorana-like}
    \end{subfigure}
    \caption{HNL oscillations in three different regimes ($\Gamma$ is the total HNL width).}
    \label{fig:oscillations}
\end{figure}

In addition, nearly degenerate HNLs can undergo coherent oscillations~\cite{Tastet:2019nqj}, i.e.\ a periodic modulation of their decay rate (with opposite phases for LNC and LNV processes) as a function of the proper time $\tau = \sqrt{(\smash[b]{x_{\mathrm{decay}}-x_{\mathrm{prod}}})^2}$ between the HNL production and decay, with the oscillation (angular) frequency given by the mass splitting $\delta M$ between the two mass eigenstates (as represented in \cref{fig:oscillations}).
We will focus on the two extreme cases:
\begin{inparaenum}[\it (a)]
    \item Dirac-like HNLs (observed before the onset of oscillations, see \cref{fig:osc_Dirac-like}) for which the rate of LNV processes is suppressed compared to LNC, and
    \item Majorana-like HNLs (observed after many oscillations, see \cref{fig:osc_Majorana-like}) for which the integrated rates\footnote{The differential distribution of the decay products will differ between LNC and LNV due to spin correlations \cite{Tastet:2019nqj}.} of LNC and LNV processes are the same.
\end{inparaenum}
Since the rates of LNC processes vanish under the single-flavor assumption, the original analysis was only sensitive to Majorana-like HNLs.

\section{Findings from the reinterpretation}
\label{sec:reinterpretation}

\begin{figure}
    \centering
    \includegraphics[width=\textwidth]{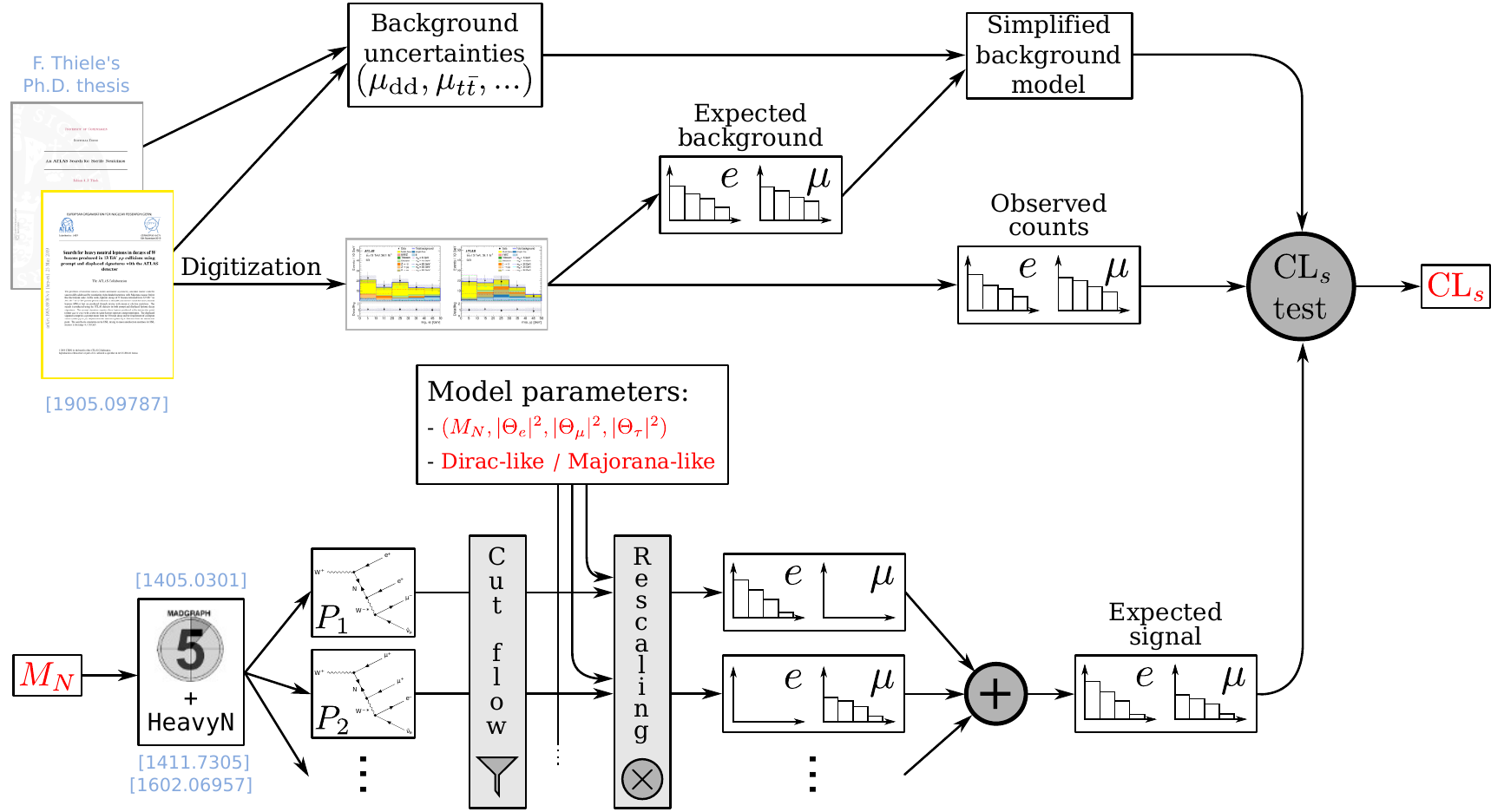}
    \caption{Sketch of the reinterpretation workflow (input and output parameters are in red).}
    \label{fig:sketch}
\end{figure}

Our reinterpretation method is described in details in ref.~\cite{Tastet:2021vwp}. \Cref{fig:sketch} attempts to briefly summarize its main features. We vary the HNL mass and mixing angles, and solve for $\mathrm{CL}_s = 0.05$ (using a simplified background model) in order to obtain the recast exclusion limit. We perform a scan for each neutrino mass hierarchy and for both Dirac- and Majorana-like HNLs. To more easily visualize the scan over the mixing angles, we define a number of \emph{benchmark points} (visible in \cref{fig:ternary_plot}) which represent both typical and extreme ratios of the squared mixing angles. In order to consistently compare different benchmarks, we express the recast limits in terms of the total mixing angle $U_{\mathrm{tot}}^2$ (summed over all three flavors and the two mass eigenstates). We finally compute a conservative bound by marginalizing over all the allowed ratios of squared mixing angles.

\paragraph{Majorana-like HNLs}

\begin{figure}
    \begin{subfigure}[b]{0.5\textwidth}
        \centering
        \includegraphics[width=\linewidth]{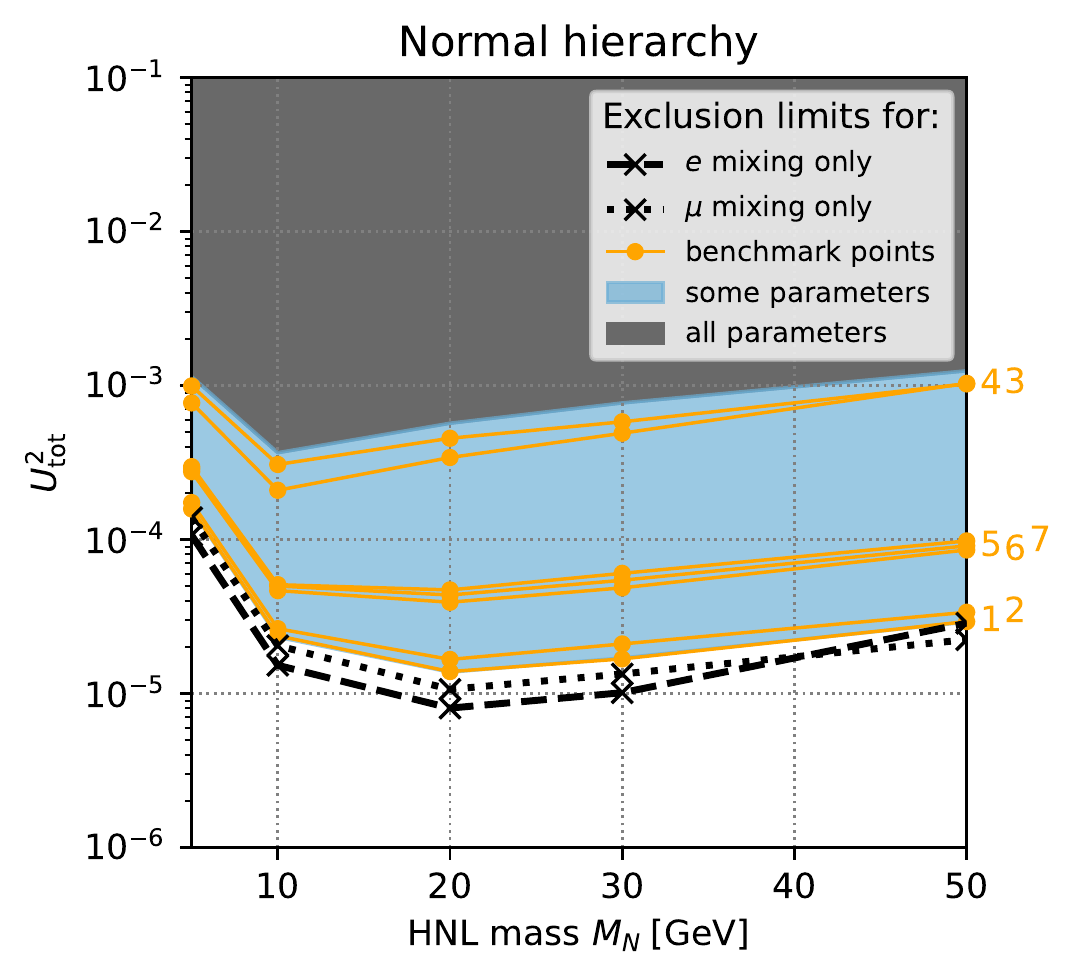}
        \caption{Majorana-like}
        \label{fig:results_Majorana}
    \end{subfigure}\hfill%
    \begin{subfigure}[b]{0.5\textwidth}
        \centering
        \includegraphics[width=0.95\linewidth]{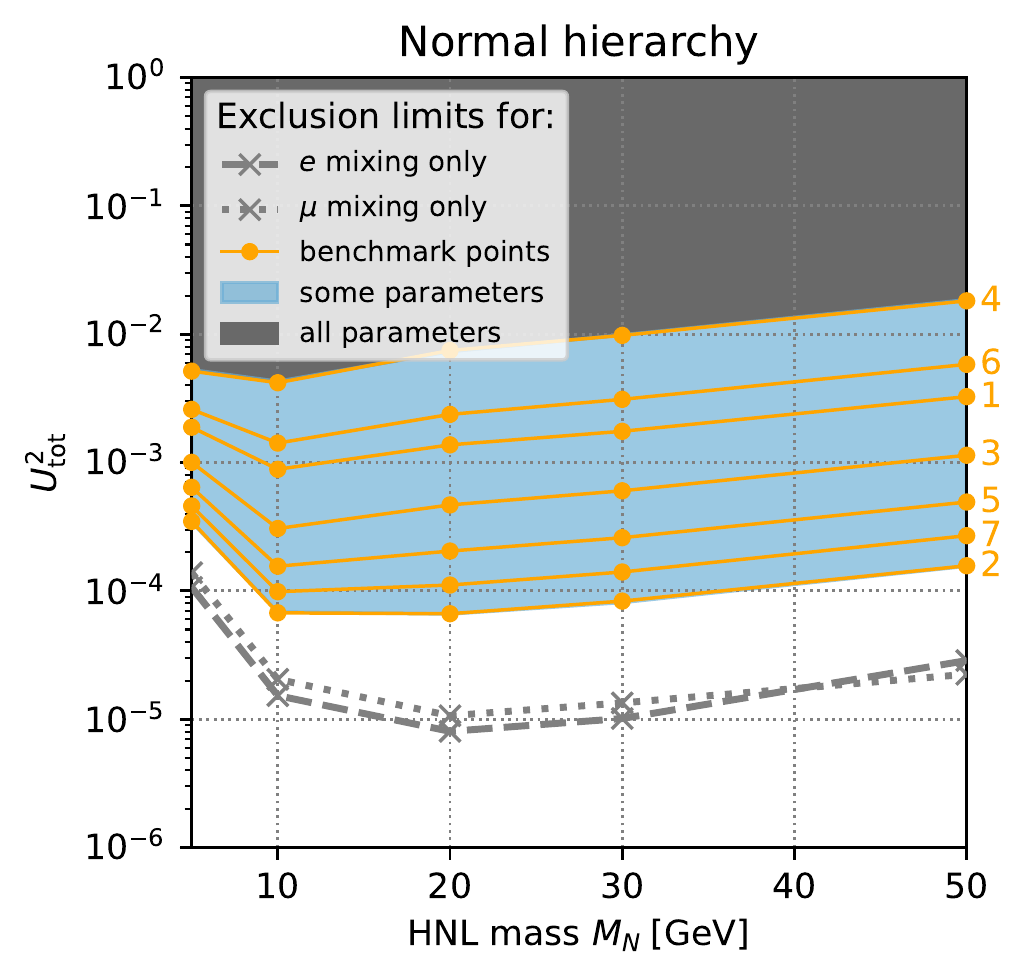}
        \caption{Dirac-like}
        \label{fig:results_Dirac}
    \end{subfigure}
    \caption{Recast limits (taking the NH as an example, the IH is similar; mind the different $y$-axes).}
\end{figure}

We obtain the recast limits shown in \cref{fig:results_Majorana}, expressed as a function of the HNL mass $M_N$ and its total mixing angle $U_{\mathrm{tot}}^2$. The black lines correspond to the simplified models originally used by ATLAS (recomputed for consistency), while the numbered colored lines denote the recast limits for the various benchmarks. We see that the recast limits can be more than an order of magnitude weaker than the original ones, with the worst case corresponding to tau flavor dominance (where the branching ratios into channels involving $\tau$ leptons are increased at the expense of the two search channels, as already observed in ref.~\cite{Abada:2018sfh} in the context of displaced searches). The blue area shows the lower and upper bounds for the recast limits when scanning over all the allowed mixing angles, and by extension the gray area is conservatively excluded within this two-HNL model.

\paragraph{Dirac-like HNLs}

The recast limits are shown in \cref{fig:results_Dirac}. Unlike in the single-flavor case where all LNC cross-sections vanish, we can now set a conservative limit thanks to the constraints from neutrino oscillations, which forbid trivial ratios of the mixing angles within this model. However, for all benchmarks, the recast limits are weaker than those obtained for a single Majorana HNL mixing with a single flavor (gray lines), by up to three orders of magnitude. The weakest limits are obtained when the electron or muon mixing angle is much smaller than the two others.

\section{Lessons learned \& recommendations for experiments}
\label{sec:conclusion}

These results show why the limits reported for simplified benchmarks \emph{should not be used directly} (e.g. by equating the $U_{\mathrm{tot}}^2$) to experimentally test more realistic models. Instead, they must be \emph{reinterpreted} within those models. Otherwise, we incur the risk of \emph{wrongly excluding} valid models or regions in parameter space.

Performing an accurate reinterpretation is a non-trivial task. In particular, computing the signal efficiencies and modeling the background can be difficult, even with a good knowledge of the experiment. To help with the former, we propose in ref.~\cite{Tastet:2021vwp} a reweighting method that allows one to exactly extrapolate the expected signal to any combination of mixing angles, using only a handful of constants that can be easily computed (and published) by experiments.
A similar method could easily be devised for other models of feebly interacting particles.

Finally, since accurately modeling the background is extremely difficult --- if not impossible --- for people working outside the experimental collaboration, a pragmatic solution that would allow theorists to reinterpret the results within their favorite model or set of parameters would be to release either
\begin{inparaenum}[\it (a)]
    \item the full likelihood, as working code, or
    \item a simplified likelihood or
    \item the covariance matrix between all background bin counts in all channels.
\end{inparaenum}
This is in line with the recommendations from the LHC Reinterpretation Forum~\cite{Abdallah:2020pec}.

\paragraph{Acknowledgments}

This project has received funding from the European Research Council (ERC) under the European Union's Horizon 2020 research and innovation programme (GA 336581, 694896) and under the Marie Skolodowska-Curie grant agreement No 860881-HIDDeN; from the Carlsberg foundation; from the Swiss National Science Foundation Excellence under grant 200020B\underline{ }182864; from the Spanish MINECO through the Centro de excelencia Severo Ochoa Program under grant SEV-2016-0597; and from the Spanish “Agencia Estatal de Investigacíon”(AEI) and the EU “Fondo Europeo de Desarrollo Regional” (FEDER) through the project PID2019-108892RB-I00/AEI/10.13039/501100011033.

\bibliographystyle{JHEP}
\bibliography{bibliography}

\end{document}